%
\magnification\magstep1
\font\cs=cmr10 scaled \magstep3
\vglue 1cm
\centerline{\cs Cosmological Perturbations seeded by}
\vskip 0.7 true cm
\centerline{\cs   Topological Defects:}
\vskip 0.7 true cm
\centerline{\cs Setting the Initial Conditions}
\vskip 1 true cm
\centerline{Nathalie Deruelle$^{1,2}$, David Langlois$^1$ and 
Jean-Philippe Uzan$^1$}
\vskip 1 true cm
\centerline{$^1$ D\'epartement d'Astrophysique Relativiste et de
Cosmologie, }
\centerline{UPR 176 du Centre National de la Recherche Scientifique,}
\centerline{Observatoire de Paris, 92195 Meudon, France}
\bigskip
\centerline{$^2$ DAMTP, University of Cambridge,}
\centerline{Silver Street, Cambridge, CB3 9EW, England}

\vskip 1cm

\centerline{July 9th, 1997}

\vskip 1cm
{\bf Pacs Numbers~:~}98.80.Cq, 98.70.Vc 
\vskip1cm

\noindent
{\bf Abstract}
\bigskip
We consider a perfectly homogeneous and isotropic universe which
undergoes a sudden phase transition. If the transition produces
topological defects, which we assume, perturbations in the geometry
and the cosmic fluid also suddenly appear. We apply the standard
general relativistic junction conditions to match the pre- and post-
transition eras and thus set the initial conditions for the
perturbations. We solve their evolution equations analytically in the
case when the defects act as a coherent source and their density
scales like the background density. We show that isocurvature as well
as adiabatic perturbations are created, in a ratio which is
independent of the detailed properties of the defects.  We compare our
result to the initial conditions currently used in the literature and
show how the cosmic fluid naturally ``compensates" for the presence of
the defects.
\vfill\eject

\font\cs=cmr10 scaled \magstep2
\noindent
{\cs 1. Introduction}
\bigskip
The question of the generation of the small cosmological
inhomogeneities at the origin of the large scale structures observed
in the universe is still open. Two main approaches, each with its
troop of specific models, are currently in competition. On one hand
the inflationary scenario (see e.g. [1]) explains cosmological inhomogeneities
by the amplification, due to accelerated cosmic expansion, of inescapable
quantum fluctuations. On the other hand, the topological defect
scenario, which is based on the idea of spatially differentiated
spontaneous symmetry breaking (see e.g. [2]), explains these inhomogeneities
by the appearance of topological defects which drive fluctuations in other
types of matter. Future observations of the small scale anisotropies
of the CMBR (Cosmic Microwave Background Radiation), in particular by
the planned MAP and PLANCK satellite missions, should discriminate
between the two scenarios.

In the present work, we shall be interested in the second one. Whereas
the calculation of cosmological perturbations in inflationary models,
at least the simplest ones, seems to have now reached a stage of
maturity and clarity, there is still some confusion in the literature
on perturbations seeded by topological defects, even on the question
of how to set the initial conditions. By initial conditions, we mean,
as is usual in cosmology, the state of cosmological perturbations for
the various matter species at a past epoch during the radiation era
when all scales of cosmological interest today were larger than the
Hubble radius. Although far in our past, this initial epoch is taken
in general very long after the phase transition supposed to have given
birth to the topological defects. The reason is that it is difficult
to trace numerically the evolution of perturbations on a very long
duration. The problem is thus to translate the ``starting" conditions
imposed by the phase transition into ``initial" conditions at the time the
numerical computations begin.

At this ``initial" time, the topological defects are thus supposed to
already exist. The main difficulty is then to set the initial value of
the perturbations of the cosmological fluids which are compatible with
the distribution of the defects. Several methods have been developped
to determine and then implement these compatibility conditions:
``integral constraints", ``compensation",
``pseudotensor"... Unfortunately, these methods are rather intricate
and confusing...

\vskip 0.5cm
The purpose of the present paper is to present a self-contained, and
hopefully clear, analytical derivation for the setting of these
initial conditions. To do so, we start our analysis before the phase
transition, when the universe is remarkably simple since it is
supposed to be strictly homogeneous and isotropic. Starting from this
pre-transition state, one should, in principle, study the detailed
evolution of the scalar field at the origin of the defects. An important
simplification arises from the  fact  that the defects can be considered as
perturbations and their evolution be obtained by solving their equations of
motion in the background geometry. However the problem remains very
complicated and requires heavy numerics. What will be retained for our
purpose is that the quantities describing the defects can be seen as
``external" sources for the evolution equations of the perturbations of the
cosmic fluids.

In order then to relate the pre-transition Friedmann Lemaitre
Robertson Walker (FLRW) spacetime to the subsequent perturbed universe
containing defects, we shall assume a sudden phase transition 
where the defects are instantaneously ``turned on".  Einstein
equations then impose some matching conditions  between the
perfectly homogeneous spacetime and the defect populated one. These
matching conditions provide constraints but are not sufficient in
themselves to determine completely the subsequent state of the
perturbations.

If one wishes to say more about the post-transition perturbations, one
must make some assumptions about the defects. The emergence
of defects can be seen as a random process and the information on
defects will thus be of a statistical nature. We shall assume that the
sources satisfy the properties of {\it causality}, {\it scaling} and 
{\it coherence}. Causality is physically required and simply states
that the sources are uncorrelated on scales larger than the Hubble
radius.  Scaling means that the statistical properties are invariant
in time up to a rescaling with respect to the Hubble radius. This
property is a simplifying one but can be justified by the convergence
towards scaling observed in numerical simulations. Here, we
assume scaling immediately (that is within one Hubble time) after the
phase transition.  Finally, coherence is a very stringent assumption,
because it means that all the statistical properties of the sources
are reduced to their statistics at a given time, the time evolution
being completely deterministic.  However, to consider coherent sources
is more general than it seems because any source for our purpose can
be decomposed into a sum of coherent ones.

The advantage of considering coherent sources is that, knowing the
time dependence of their correlators from scaling and causality, one
can then solve for the evolution of the perturbations by a
simple use of Einstein's equations.  In the long wavelength limit,
i.e. for scales larger than the Hubble radius, one obtains the
solution explicitely as a sum of power-law terms.

From our explicit analytical solutions, we are able to justify some of
the statements which can be found in the literature. In particular,
the solutions for the perturbations, which can be decomposed into
terms driven by the sources and ``homogeneous" (in the sense of
differential equations) terms, are shown to be dominated by the source
driven terms.  We also recover on our solutions the so-called
phenomenon of ``compensation".  Another conclusion of our work, which
is original to our knowledge, is that, when one allows for cold dark
matter, the sources will drive the long-wavelength perturbations into
a combination of adiabatic and isocurvature perturbations, the
relative ratio being a universal constant.

\vskip 0.5cm

The paper is organized as follows.  In section 2, we write the
linearised Einstein equations.  Our formalism is based on the usual
cosmological perturbation theory using Bardeen type gauge-invariant
quantities, where one distinguishes between scalar, vector and tensor
perturbations.  In addition to the usual geometrical and perfect fluid
type matter variables, one must introduce the energy momentum tensor
of the defects.  This tensor is automatically gauge invariant because,
as stated before, the defects are considered as perturbations.  In
Section 3, we describe the phase transition. In Section 4 we review
the matching conditions on a constant energy density surface and apply
them to the case of a sudden phase transition in the universe. They
enable us to make the link between the unperturbed universe and the
post-phase transition perturbed universe, or rather they give
constraints that physically admissible configurations must satisfy.
Section 5 deals with the statistical properties of the sources and the
notions of coherence, scaling and causality are introduced in detail.
In section 6, we give the behaviour of perturbations larger than the
Hubble radius after the transition, which will constitute the
``initial" conditions. Finally, section 7 comments on the obtained
solutions.

\bigskip

\noindent
{\cs 2. Linearised Einstein equations}
\bigskip
{\bf 2.1 The background}
\bigskip
The universe at large appears to be remarkably homogeneous and
isotropic and governed by the gravitational force created by its
material content, to wit a mixture of radiation and dust. It is
therefore well described by a Robertson-Walker geometry whose time
evolution satisfies Friedmann's equations. Phase transitions [3] [2] can
have occured in the very early universe when it was pure radiation and
its spatial curvature negligible. We thus take the background line
element in that era to be~:
$$ds^2=a^2(\eta)(-d\eta^2+\delta_{ij}dx^idx^j)\eqno(1)$$ where
$x^0\equiv\eta$ is conformal time, $x^i$, $i=1,2,3$, three
cartesian coordinates, and $a(\eta)$ the scale factor. The Friedmann
equations are~: $$ {\cal H}'=-{\cal H}^2\qquad ,\qquad\kappa\rho
a^2=3{\cal H}^2\eqno(2)$$ where a prime denotes a derivative with
respect to conformal time, where ${\cal H}\equiv a'/ a$, where $\rho$
is the radiation energy density and $\kappa\equiv8\pi G$ is Einstein's
constant. The solution of (2) is~: $${\cal
H}={1\over\eta}\quad\Rightarrow\quad a=a_0\eta\quad ,\quad\kappa\rho
a^2={3
\over\eta^2}\quad\Rightarrow\quad{\rho'\over\rho}=-{4\over\eta}
\eqno(3)$$ 
where $a_0$ is an integration constant which can be chosen so
that
$a=1$ today.
\bigskip
{\bf 2.2 Scalar perturbations}
\bigskip
Following Bardeen [4] we split the perturbations of the geometry and
the matter variables into ``scalar", ``vector" and ``tensor" parts. In
this paragraph we define the scalar part and write the linearised
Einstein equations for the gauge invariant scalar perturbations (for
reviews of this formalism, see e.g.  [5-7]).

The line element of a perturbed Robertson-Walker space time reads,
when the perturbations are scalar and the background given by (1-3)~:
$$ds^2=a_0^2\eta^2\left[-(1+2A)d\eta^2+2\partial_iBdx^id\eta+\left\lbrace
(1+2C)\delta_{ij}+2\partial_{ij}E\right\rbrace
dx^idx^j\right]\eqno(4)$$ where $A, B, C, E$ are four ``small"
functions of space and time.  Under an infinitesimal coordinate
transformation $$\eta\rightarrow\eta+T\qquad,\qquad x^k\rightarrow
x^k+\partial^kL\eqno(5)$$ where $T$ and $L$ are two arbitrary first
order functions of $\eta$ and $x^i$, the four scalar metric
perturbations $A,B,C,E$ transform as $$A\rightarrow
A+T'+{T\over\eta},\quad B\rightarrow B-T+L',\quad C\rightarrow
C+{T\over\eta},\quad E\rightarrow E+L.\eqno(6)$$ One can thus
introduce two gauge independent scalar metric perturbations, for
example~:
$$\Psi=-C-{1\over\eta}(B-E')\qquad,\qquad\Phi=A+{1\over\eta}(B-E')
+(B-E')'. \eqno(7)$$

\vskip 0.5cm
The energy-momentum tensor of the matter content of this perturbed
universe can be written as~: $$T_{\mu\nu}=\bar T_{\mu\nu}+\delta
T_{\mu\nu}+\Theta_{\mu\nu}.\eqno(8)$$ $\bar T_{\mu\nu}$
($\mu,\nu=0,1,2,3$) is the energy-momentum tensor of the homogeneous
and isotropic radiation background~; $\delta T_{\mu\nu}$ is its
perturbation~: its scalar components can be expressed in terms of two
scalar quantities, $\delta\equiv\delta\rho/\rho$, the density
contrast, and $v$, the velocity perturbation, as (see e.g. [5-7])~:
$$\eqalign{\kappa\delta T^S_{00}&={3\over\eta^2}(\delta+2A)\quad,\quad
\kappa\delta
T^S_{0i}=-{3\over\eta^2}\partial_i\left(B+{4\over3}v\right)\cr
\kappa\delta
T^S_{ij}&={1\over\eta^2}\left[\delta_{ij}(2C+\delta)+2\partial_{ij}E\right]
.\cr}\eqno(9)$$
(Note that we describe the material content of the universe as a
single radiation fluid (with no anisotropic stresses). This is
justified since at the era of the phase transition all matter is
highly relativistic.)

In the coordinate transformation (5) $\delta$ and $v$ transform as~:
$\delta\to\delta-4T/\eta$, $v\to v-L'$, so that two gauge invariant
scalar perturbations for the radiation fluid can be constructed,
e.g.~: $$\delta^\flat=\delta+4C\qquad,\qquad
v^\natural=v+E'.\eqno(10)$$ Instead of $\delta^\flat$ one can also
use~: $$\delta^\natural=\delta-{4\over\eta}(B+v)\quad{\rm or}\quad
\delta^{\sharp}=\delta-{4\over\eta}(B-E').\eqno(11)$$
$\delta^\natural$, $\delta^{\sharp}$ and $\delta^\flat$ are the density
contrasts which are respectively defined in the comoving gauge where
$\delta T^0_i=0$, the newtonian (or longitudinal) gauge, and in the
flat slicing gauge. They are related by
$$\delta^\natural=\delta^{\sharp}-{4\over\eta}v^\natural\quad {\rm and}\quad
\delta^\flat=\delta^{\sharp}-4\Psi. \eqno(12)$$

Finally $\Theta_{\mu\nu}$ is the energy-momentum tensor of the scalar
field at the origin of the topological defects. We suppose that it is
a small perturbation which does not contribute to the background (this
is the so-called ``stiff approximation" (see e.g. [8])). We decompose its
scalar components as~:
$$\Theta^S_{00}=\rho^s,\quad \Theta^S_{0i}=-\partial_i v^s,\quad
\Theta^S_{ij}=\delta_{ij}\left(P^s-{1\over3}\Delta\Pi^s\right)+
\partial_{ij}\Pi^s. 
\eqno(13)$$
The four source functions $\rho^s, P^s, v^s, \Pi^s$ will be discussed
later.  They are gauge invariant since $\Theta_{\mu\nu}$ is a tensor
which vanishes in the unperturbed background.
\vskip 0.5cm

Having defined gauge invariant scalar perturbations for the metric (eq
(7)), for the radiation fluid (eq (10-12)) and for the sources (eq
(13)), we now write their evolution equations.  We shall write them in
Fourier space, the Fourier transform of any function $f(x^i,\eta)$
being defined as~ $$\hat f(k^i,\eta)\equiv{1\over(2\pi)^{3/2}}\int\!
d^3\!x\, e^{-{\rm i}k_ix^i}f(x^i,\eta)\quad
\Leftrightarrow \quad f(x^i,\eta)={1\over(2\pi)^{3/2}}\int\!
 d^3\!k\,e^{{\rm i}k_ix^i}\hat
f(k^i,\eta).\eqno(14)$$

The conservation equations for the radiation fluid, when the
background is governed by equations (2-3), can be cast under the form
(see ref [5-7])~: $$\hat\delta^{\flat '}={4\over3}k^2\hat
v^\natural\eqno(15a)$$ $$k_i\left(\hat v^{\natural '}+\hat
\Phi+{1\over4}\hat\delta^{\sharp}\right)=0.\eqno(15b)$$ The conservation
equations for $\Theta_{\mu\nu}$ read~:
$$\hat\rho^{s'}+{1\over\eta}(\hat\rho^s+3\hat P^s)-k^2\hat
v^s=0\eqno(15c)$$ $$k_i\left(\hat v^{s'}+{2\over\eta} \hat v^s+
\hat P^s-{2\over3}k^2\hat\Pi^s\right)=0.\eqno(15d)$$ 
For all but the $k=0$ mode, the linearised Einstein equations read~:
$$\hat\Psi-\hat\Phi=\kappa\hat\Pi^s\eqno(15e)$$
$$-k^2\hat\Psi={3\over2\eta^2}\hat\delta^\natural
+{\kappa\over2}\left(\hat\rho^s-{3\over\eta}\hat v^s\right)\eqno(15f)$$ 
$$\hat\Psi'+{1\over\eta}\hat\Phi=-{2\over\eta^2}\hat
v^\natural-{\kappa\over2}\hat v^s\eqno(15g)$$
$$\hat\Psi''+{4\over\eta}\hat\Psi'+{1\over3}k^2\hat\Psi=\kappa
\left(-{1\over3}k^2\hat\Pi^s+
{\hat\Pi^{s'}\over\eta} +{1\over2}\hat P^s-{1\over6}\hat\rho^s\right).
\eqno(15h)$$

In section 6 we shall solve this set of eight equations for the four
unknowns $\hat\Psi,
\hat\Phi,\hat\delta^\natural$ and $\hat v^\natural$~: the source functions
$\hat\rho^s, \hat P^s, \hat v^s,
\hat\Pi^s$, subject to the constraints (15c) and (15d)
 being known, eq (15h) will give the
metric perturbation $\hat\Psi$. Then $\hat\Phi$ is given by (15e),
$\hat\delta^\natural$ by (15f) and $\hat v^\natural$ by (15g). The
last two equations, (15a) and (15b) are redundant (Bianchi identity)
and can be used as a check of the calculation.
\bigskip
{\bf 2.3 Vector perturbations}
\bigskip
 The line element  for ``vector" perturbations reads
$$
ds^2=a_0^2\eta^2\left[-d\eta^2+2{\bar B}_idx^id\eta+\left\lbrace
\delta_{ij}+2\partial_{(i}{\bar E}_{j)}\right\rbrace dx^idx^j
\right],\eqno(16)$$
where $\bar B^i$ and $\bar E^i$ are small functions of space and time
subject to the condition~: $\partial_i{\bar B}^i=\partial_i{\bar
E}^i=0$. (Henceforth all ``barred" quantities $\bar V^i$ will be
divergenceless vectors~: $\partial_i \bar V^i=0$.) Under the
infinitesimal coordinate transformation 
$$\eta\to\eta\qquad,\qquad
x^i\to x^i+\bar L^i\eqno(17)$$ 
where $\bar L^i$ is an arbitrary first
order divergenceless vector, the four components of the two vector
perturbations $\bar B_i$ and $\bar E_i$ transform as $$\bar B_i\to
\bar B_i+\bar L'_i\quad,\quad\bar E_i\to\bar E_i+\bar L_i,\eqno(18)$$
so that the two components of the vector perturbations $$\bar \Phi_i=
\bar E_i'-\bar B_i\eqno(19)$$ are gauge invariant.

The non-zero vector components of $\delta T_{\mu\nu}$, the
perturbation of the energy-momentum tensor of the radiation fluid, can
be expressed in terms of $\bar v^i$, the vector perturbation of the
fluid velocity, as (see ref [5-7]) $$\kappa\delta
T^V_{0i}=-{1\over\eta^2}(3\bar B_i+4\bar v_i),\quad
\kappa\delta T^V_{il}={2\over\eta^2}\partial_{(i}\bar E_{l)}.\eqno(20)$$
In a coordinate transformation~: $\bar v^i\to \bar v^i-\bar
L^{i\prime}$, so that $$\bar v_i^{\sharp}=\bar v_i+\bar B_i\eqno(21)$$
is a gauge invariant quantity.

As for the non-zero vector components of the energy-momentum tensor of
the scalar field at the origin of the topological defects we write
them as $$\Theta^V_{0i}=-\bar v^s_i\quad{\rm and}\quad
\Theta^V_{ij}=2\partial_{(i} {\bar\Pi}^s_{j)}\eqno(22)$$ where the
four source functions $\bar v^s_i$ and $\bar\Pi^s_i$ are gauge
invariant and will be discussed in section 5.
\vskip 0.5cm

The gauge invariant vector perturbations for the metric (eq (19)), for
the radiation fluid (eq (21)) and for the sources (eq (22)) having
thus being defined, we now write their evolution equations.

The equation of conservation for the radiation fluid is (in Fourier
space) $$\hat{\bar v}_i^{{\sharp}\prime}=0,\eqno(23a)$$ and is $$\hat{\bar
v}_i^{s\prime}+{2\over\eta}\hat {\bar
v}^s_i-k^2\hat{\bar\Pi}^s_i=0\eqno(23b)$$ for the defects.  The
Einstein equations split into
$$-k^2\hat{\bar\Phi}_i=-{8\over\eta^2}\hat{\bar v}_i^{\sharp}-2\kappa
\hat{\bar v}^s_i\eqno(23c)$$
$$\hat{\bar\Phi}_i'+{2\over\eta}\hat{\bar \Phi}_i=2\kappa\hat{\bar\Pi}^s_i.
\eqno(23d)$$

The four source functions $\hat{\bar v}^s_i$ and $\hat{\bar\Pi}^s_i$
being known and subject to the constraint (23b), eq (23d) and (23c)
will yield the metric perturbation $\hat{\bar\Phi}_i$ and the velocity
perturbation $\hat{\bar v}_i$. Eq (23a) is redundant.

\bigskip
{\bf 2.4 Tensor perturbations}
\bigskip
The tensor perturbations of the geometry are defined by
$$
ds^2=a_0^2\eta^2[-d\eta^2+(
\delta_{ij}+2{\bar E}_{ij}) dx^idx^j],
\eqno(24)$$
with $\partial_i\bar E^{ij}=\bar E_i^i=0$. ${\bar E}_{ij}$ is gauge
invariant. The non-zero tensorial components of the perturbation of
the energy-momentum tensor of the radiation fluid reduce to~: $$\kappa
\delta T^T_{ij}={2\over3\eta^2}\bar E_{ij}\eqno(25)$$ and the evolution of
${\bar E}_{ij}$ is given in Fourier space by (see ref [5-7])
$$\hat{\bar E}_{il}''+{2\over\eta}\hat{\bar E}_{il}'+k^2\hat{\bar
E}_{il}= 2\kappa\hat{\bar\Pi}^s_{il},\eqno(26)$$ where
$\hat{\bar\Pi}^s_{il}$ is the tensorial part of the energy-momentum
tensor of the scalar field giving rise to the topological defects.
\bigskip
\noindent
{\cs 3. The phase transition}
\bigskip

Topological defects are formed in the very early universe during a
phase transition when some scalar field acquires different vacuum
expectation values in different regions of space [2] [3].

Before the phase transition, the field is zero everywhere and its
``false vacuum" potential energy is constant. Its energy-momentum
tensor $\Theta_{\mu\nu}$ can then be modelled by a small cosmological
constant. Hence, the only non vanishing source functions are
$$\rho^s=\lambda\eta^2,\quad P^s=-\lambda \eta^2,\eqno(27)$$ where
$\lambda$ is a positive constant related to the false vacuum
energy. They satisfy the constraint (15c) (eq (15d) is empty since
$P^s$ does not depend on the space coordinates).

 Before the phase transition then, the universe is strictly
 homogeneous and isotropic. Hence the separation made here between the
 background governed by the radiation fluid and the perturbations
 caused by the scalar field in its false vacuum state is
 artificial. Indeed a strictly homogeneous and isotropic perturbation
 can always be absorbed into a redefinition of the background
 scale factor. We introduce it however for the sake of
 clarity.
\vskip 0.5cm

As the temperature of the radiation fluid drops below a given critical
temperature the field rolls down or tunnels to a ``true vacuum" state
of zero energy.  Bubbles of true vacuum are formed, grow and
percolate. If the true vacuum manifold is degenerate topological
defects appear at the boundaries of bubbles characterized by different
true vacuum values of the field [2-3].

The precise energy-momentum tensor of the field after this phase
transition depends on the nature of the defects formed and can be
obtained solely by means of heavy numerical calculations (see e.g. [9-11]).
However some of its statistical properties can be inferred from general
arguments (see below Section 5).

The duration $\Delta\eta$ of the phase transition itself must be brief
(if it is delayed for too long the universe may enter an inflationary
phase [12] [1], and we assume that this does not happen).  In fact we assume
that it is less or of the order of one Hubble time $$\Delta\eta\simeq
1/{\cal H}\quad \hbox{with}\quad {\cal H}={1\over\eta_{PT}}\eqno(28)$$ where
$\eta_{PT}$ is the epoch of the phase transition.  Hence the phase
transition will look instantaneous to perturbations evolving on time
scales greater than $\Delta\eta$ that is to modes $k$ such that
$$1/k\gg\Delta\eta\quad\Leftrightarrow\quad k\eta_{PT}\ll1\eqno(29)$$
that is to modes which are larger than the horizon at the epoch of the
transition. Now all modes of interest today were larger than the
horizon in the early universe. It is therefore justified to describe a
transition which lasts less than a Hubble time as instantaneous.

Thus, we only need to match the spacetime geometries and the matter
variables on the surface of transition that is the surface $\Sigma$ of
constant temperature (or constant density). This will be done in the
next section by imposing that the induced three metric on $\Sigma$ and
the extrinsic curvature of $\Sigma$ must be continuous (see e.g
[13]).

\bigskip
\noindent
{\cs 4. Matching conditions on a constant energy density surface}
\bigskip
Following [14], we first write the matching
conditions in a gauge where $\Sigma$ is a constant time hypersurface
and then translate them in an arbitrary coordinate system. (The
difference with [14] is that here the background geometry evolves
smoothly during the transition so that ${\cal H}$ and ${\cal H}'$ are
continuous.)
\bigskip
{\bf 4.1 Scalar modes}
\bigskip
In an arbitrary coordinate system in which the line element is given
by (4) the surface of transition $\Sigma$ is defined by
$q(\eta,x^k)\equiv q_0(\eta)+\delta q=Const$ where $q$ is a scalar (we
shall take $q$ to be the total energy density). In the coordinate
transformation (5) $\delta q\to\delta q+Tq'_0$. Choosing $T=-\delta
q/q'_0$ (and $L$ arbitrary) therefore takes us to the gauge where the
surface $\Sigma$ is a constant time surface. In that gauge, that we
shall label with a tilde, the unit normal vector to the constant time
hypersurfaces is $$n_0=-a(1+\tilde A)\quad{\rm and}\quad
n_i=0.\eqno(30)$$ The induced metric and the extrinsic curvature are
defined by $$\perp_{\mu\nu}=g_{\mu\nu}+n_\mu n_\nu,\quad{\rm and}\quad
K_{\mu\nu}=-{1\over2}{\cal L}_n \perp_{\mu\nu},\eqno(31)$$ where
${\cal L}$ stands for the Lie derivative. Their non vanishing scalar
components are given by $$\eqalign{
\perp_{ij}&=a_0^2\eta^2[(1+2\tilde C)\delta_{ij}+2\partial_i\partial_j\tilde E]
\cr
K^i_j&=-{1\over a_0\eta}\left[{1\over\eta}\delta^i_j+ 
\left(-{1\over\eta}\tilde A+\tilde C'\right)\delta^i_j-
\partial^i\partial_j(\tilde B-\tilde E')\right].\cr}\eqno(32)$$

At first order, when one shifts back to the original, arbitrary gauge
(using (6)), the continuity of $\perp_{\mu\nu}$ and $K_{\mu\nu}$ then
imposes $$\eqalign{
\left[C-{1\over\eta_{PT}}{\delta q\over q_0'}\right]_\pm&=0 \quad,\quad
\left[\partial^i\partial_j(E+L)\right]_\pm=0\cr
\left[\partial^i\partial_j\left(E'-B-{\delta q\over q_0'}
\right)\right]_\pm&=0\quad,\quad
\left[-{1\over\eta_{PT}}A+C'+{2\over\eta_{PT}^2}{\delta q\over q_0'}
\right]_\pm=0,
\cr}\eqno(33)$$
where $[F]_\pm$ is defined as $[F]_\pm={\rm lim}_{\epsilon\rightarrow 0^+}
\left[F(\eta_{PT}+\epsilon)-F(\eta_{PT}-\epsilon)\right]$ and 
where $\eta_{PT}$ is the conformal time at which the transition occurs
and $q$ is the surface of constant total density, so that
$$q'_0=\rho'\quad,\quad \delta q=\rho\delta+\rho^s/a^2.\eqno(34)$$ The
second condition in (33) is empty since one can always choose $L$ such
that it is fulfilled.  As for the other three, they can be rewritten
in terms of scalar gauge invariant quantities and in Fourier space, as
$$\eqalign{\left[{3\over\eta_{PT}^2}\hat \delta^\flat+\kappa\hat
\rho^s\right]_\pm=0,\quad
\left[k^ik_j\left({3\over\eta_{PT}^2}\hat \delta^{\sharp}+\kappa\hat
 \rho^s\right)
\right]_\pm=0\cr
\left[{1\over\eta_{PT}}\hat \Phi+\hat \Psi'+{1\over2\eta_{PT}}\left(\hat 
\delta^{\sharp}+{1\over3}\kappa\hat
\rho^s\eta_{PT}^2\right)\right]_\pm=0\cr}\eqno(35)$$ and it is an easy
exercise (which makes use of the linearised equations (15)) to show
that the third is redundant.

To summarize, the two independent matching conditions for the scalar
perturbations can be simply written as (using the relation (12)
between $\delta^\flat$ and $\delta^{\sharp}$) $$\left[{3\over\eta^2_{PT}}\hat
\delta^\flat+\kappa\hat \rho^s\right]_\pm=0\quad,\quad
\left[k^ik_j\hat\Psi\right]_\pm=0.\eqno(36)$$

\bigskip
{\bf 4.2 Vector and tensor modes}
\bigskip

The normal vector to the constant time hypersurfaces does not have any
vector nor tensor component and thus, $$n_0=-a\quad{\rm and}\quad
n_i=0.\eqno(37)$$ This leads to the following expression for the
vector components of the induced three metric and the extrinsic
curvature,
$$\perp_{ij}=a_0^2\eta^2[\delta_{ij}+2\partial_{(i}\bar{E}_{j)}]
\quad,\quad
\delta K^j_i={1\over2a_0\eta} (\partial_i{\bar \Phi}^j+
\partial^j{\bar \Phi}_i).\eqno(38)$$
As for their tensor components they are
$$\perp_{ij}=2a_0^2\eta^2\bar E_{ij}\quad,\quad
\delta K^i_j={1\over a_0\eta}\bar E^{'i}_j.\eqno(39)$$

The matching conditions then reduce to, in Fourier space
$$[k_{(i}\hat{\bar\Phi}_{j)}]_\pm=0\qquad\hbox{and}\qquad [\hat{\bar
E}_{ij}]_\pm=0\quad{\rm ;}\quad [\hat{\bar E}'_{ij}]_\pm =0.\eqno(40)$$

\bigskip
{\bf 4.3 The case of a phase transition}
\bigskip

In the particular case of the phase transition described in Section 3,
the perturbations are strictly homogeneous and isotropic before the
transition so that their Fourier transform is a zero mode proportional
to the Dirac distribution $\delta(k)$. After the transition, all
perturbations depend on space (apart from the zero mode which can be
absorbed into a redefinition of the background). In Fourier space then
the matching conditions are, for all modes apart the strictly $k=0$
one~: $$\eqalign{
\left[{3\over\eta_{PT}^2}\hat\delta^\flat+\kappa\hat\rho^s\right]_{\eta_{PT}}=0
\quad,\quad
\left[\hat\Psi\right]_{\eta_{PT}}=0\cr
[\hat{\bar\Phi}_i]_{\eta_{PT}}=0\quad,\quad [\hat{\bar
E}_{ij}]_{\eta_{PT}}=0\quad,\quad [\hat{\bar E}'_{ij}]_{\eta_{PT}}
=0.\cr}\eqno(41)$$ There are two conditions for the scalar modes, one
for the vector ones and two for the tensor ones, which is the required
number since the equations of evolution for $\Psi$ and ${\bar E}_{ij}$
are second order and the one for ${\bar\Phi}_i$ is first order. Note
that these conditions, because we assumed that the universe was
strictly homogeneous and isotropic before the transition, do not
depend on the physics before the transition, for example on the value
of the false vacuum energy.

What needs to be done now is to propagate the matching conditions (41)
to a much later time in the radiation era, in order to set
analytically the effective initial conditions which must be taken in
the numerical integration of the evolution equations for the
perturbations. In order to do that we need to specify the
energy-momentum tensor of the sources (Section 5) and analytically
integrate the evolution equations from the epoch of the phase
transition to the time when the numerical integration starts (Section
6).

\bigskip
\noindent
{\cs 5. The energy-momentum tensor $\Theta_{\mu\nu}$ of the sources}
\bigskip
{\bf 5.1 Coherent vs independent sources}
\bigskip

When the phase transition occurs the scalar field settles randomly in
its true vacuum state in uncorrelated spatial domains. The
distribution however is statistically homogeneous and isotropic
because the background is so and the physics of the transition is
supposed to obey the cosmological principle.

The ten components of the energy-momentum tensor
$\Theta_{\mu\nu}(\eta,x^i)$ of the topological defects created are
therefore ten random fields, as well as their ten Fourier transforms
$\hat\Theta_{\mu\nu}(\eta,k^i)$ (which are complex but such that
$\hat\Theta^*_{\mu\nu}(\eta,k^i)=\hat\Theta_{\mu\nu}(\eta,-k^i)$). From
now on, we ignore the ($k=0$) mode which can be absorbed in the
background.

The statistical properties of these ten random fields, that we shall denote
collectively by $ S_a(\eta, x^i)$ or $\hat S_a(\eta, k^i)$, are determined by
the values of their correlators. Since we have in view the computation of the
two-point correlator of the temperature anisotropies of the microwave
background, the quantities we need to know are the unequal time
two-point correlators of the sources, that is 
$$\langle S_a(\eta, x^i)
S_{a'}(\eta',x'^i)\rangle \equiv C_{a,a'}(\eta,\eta',r)\eqno(42)$$
where $\langle ...\rangle $ means an ensemble average on a large
number of different realisations of the transition, and where the
correlator $C_{a,a'}$ depends only on $r\equiv|\vec x-\vec x'|$
because of the homogeneity and isotropy of the distribution. The power
spectrum of the correlator $C_{a,a'}$ is defined as $$
P_{a,a'}(\eta,\eta',k)\equiv (2\pi)^{3/2}\hat
C_{a,a'}(\eta,\eta',k)\eqno(43)$$ where a hat denotes a Fourier
transform (see (14)) and where the dependence in $k\equiv |\vec k|$,
as well as the fact that $P_{a,a'}(\eta,\eta',k)$ is real are again
due to the homogeneity and isotropy of the distribution. The power
spectrum is related to the correlators in Fourier space by $$\langle
\hat S^*_a(\eta,k^i)\hat S_{a'}(\eta',k'^i)\rangle =\delta ( k^i-
k'^i) P_{a,a'}(\eta,\eta',k).\eqno(44)$$

 As shown by Turok [15], a clever way to look at the power spectra
 $P_{a,a'}(\eta,\eta',k)$ is to see their ensemble as a matrix $M$
 where the columns are labelled by the indices $(a,\eta)$ and the rows
 by the indices $(a',\eta')$.  Because the power spectra are real this
 matrix is symmetric.  Symmetric matrices can be diagonalised (we
 ignore here the fact that $M$ is of infinite dimension)
 so that we have 
$$P_{a,a'}(\eta,\eta',k) =
\sum_{\tilde a}\int_{\tilde\eta}\!
{d\tilde\eta}\,\,p^{\tilde a\tilde\eta}_{a\eta}(k)\,\lambda_{\tilde
a\tilde\eta}\, p^{\tilde a\tilde\eta}_{a'\eta'}(k)\eqno (45)$$ 
where $\lambda_{\tilde a\tilde\eta}$ are the infinite number of possibly
degenerate and positive eigenvalues of the matrix $M$.
\vskip 0.5cm

Depending on the numerical
value of the eigenvalues $\lambda_{\tilde a\tilde\eta}$ and the associated
(normalised) eigenvectors $p^{\tilde a\tilde\eta}_{a\eta}$, the sources can
be called coherent, independent, or mixed.

Perfect ``time coherence" of a given random field $S_a$, say, means
that its power spectrum factorizes, that is 
$$P_{ a, a}(\eta,\eta',k)= p_a(\eta,k) p_a(\eta',k).
\eqno(46)$$
(This means that in (45) $\lambda_{\tilde a\tilde\eta}$ is of the
form $\lambda_{\tilde a\tilde\eta}=\lambda\delta_{\tilde a a}
\delta(\tilde\eta-\bar\eta)$,  and
$p_a(\eta,k)\equiv \sqrt{\lambda}p^{a\bar\eta}_{a\eta}$.)
$p_a(\eta,k)$
 is the square root of the power spectrum of
the equal time auto-correlator of $S_a$~: $\langle \hat
S_a^*(\eta,k^i)\hat S_a(\eta,k'^i)
\rangle=
\delta(k^i-k'^i)\left(p_a(\eta,k)\right)^2$.
 
In contrast, perfect ``time independence" means that
$$P_{ a,
a}(\eta,\eta',k)=\delta(\eta-\eta')P_a(\eta,k).\eqno(47)$$ 
(This means that in (45) $\lambda_{\tilde 
a\tilde\eta}=\lambda\delta_{\tilde a a}$, 
$p^{a\tilde\eta}_{ a\eta}(k)=p_{
a\eta}\delta(\tilde\eta-\eta)$, and that $P_a(\eta,k)\equiv\lambda
\left(p_{
a\eta}\right)^2$.)

Textures, which evolve fairly smoothly, tend to be time coherent,
whereas local cosmic strings, which undergo complex processes of
intercommutation, tend on the other hand to be time independent (see
e.g. [16]). It is clear that the evolution of time independent sources is more
difficult to describe since at each moment a new realization of the random
fields is drawn. In the following we shall consider time coherent
sources only.

Time coherence (or independence), which concerns autocorrelators
only, does not however characterize completely the statistical
properties of the sources. Let us then turn to  cross correlators.

\vskip 0.5cm
A given subset 
 $\{S_a, a\in A\}$  of the ten sources is statistically coherent if
$$P_{a,a'}(\eta,\eta',k)= p_a(\eta,k) p_{a'}(\eta',k).\eqno (48)$$
(This means that in (45) the eigenvalues $\lambda_{\tilde a\tilde\eta}$ are 
zero for all $\tilde a$ not in the subset $(a,a')$ and all $\tilde\eta$
not equal to some $\bar\eta$.) (In keeping with the terminology used in
quantum mechanics, such
 sources $S_a$ can also be called ``pure", see
[15]). Note that statistical coherence
 of the random fields among themselves, that is property (48), implies
 time coherence, that is property (46).

Knowing the unequal time two-point correlators (42) or (44), subject
to the coherence condition (48), is not enough a priori to specify
completely the random fields $S_a(\eta,x^i)$ of the subset.  However a large
class of random fields satisfying (48) is $$\hat S_a(\eta,k^i)=
p_a(\eta,k)\,\,\, e_A(k^i)\eqno(49)$$ where $e_A(k^i)$ is a
normalized complex random variable characterising the subset $A$~:
$\langle e_A^*(k^i)e_A(k'^i)\rangle =\delta(k^i-k'^i)$.  It is clear
that as long as one computes two-point correlators only the two definitions
(48) and (49) are equivalent. (Beware of the fact that the left-hand side of
(49) is not some ``typical" realisation of the random field but the random
field itself, as should be clear from the explicit introduction of the
random variable $e_A(k^i)$  in the
right-hand side. Omitting
$e_A(k^i)$ on grounds of short-hand
notations could for example induce into thinking that $\hat S_a(\eta,k^i)$ is
a real quantity~!)

\vskip 0.5cm

Consider now two disjoint subsets, $\{S_a, a\in A\}$, $\{S_b, b\in B\}$, 
each being
coherent, that is such that $S_a$ is given by (49) and $S_b$ by a similar
expression~: $\hat S_b(\eta,k^i)=
p_b(\eta,k)\,\,\, e_B(k^i)$, $e_B(k^i)$ being the random variable
characterising the subset $B$, with
$\langle e_B^*(k^i)e_B(k'^i)\rangle =\delta(k^i-k'^i)$. 

The two subsets $A$
and $B$ are said to be statistically ``independent" if $\langle
e_A^*(k^i)e_B(k'^i)\rangle =\delta_{AB}\delta(k^i-k'^i)$. 

The difference between coherent and independent sources is most
strikingly seen from the power spectrum of the sum of two different
random fields. In the first case ($a$ and $a' \in A$)
$$\langle [\hat S^*_a(\eta,k^i)+\hat S^*_{a'}(\eta,k^i)][\hat
S_a(\eta,k'^i)+\hat
S_{a'}(\eta,k'^i)]\rangle=\left[p_a(\eta,k)+p_{a'}(\eta,k)\right]^2\delta(k^i-k'^i)
\eqno(50)$$
whereas in the second ($a\in A, b\in B$) $$\langle [\hat S^*_a(\eta,k^i)+\hat
S^*_b(\eta,k^i)][\hat S_a(\eta,k'^i)+\hat
S_b(\eta,k'^i)]\rangle=\left(\left[p_a(\eta,k)\right]^2+
\left[p_b(\eta,k)\right]^2\right)\delta(k^i-k'^i).
\eqno(51)$$

In the following we shall consider statistically independent subsets
$(A, B...)$ of statistically (and time) coherent sources ($S_a,
S_b,...)$.  Indeed, we shall see that causality imposes that the
sources be divided into statistically independent subsets and we shall
assume that within each subset the sources are coherent. This covers a
large class of sources. Indeed, according to (45), the correlators of
more complex, ``mixed" (or partially coherent) sources can be
decomposed into a sum of correlators of coherent sources.

\bigskip
{\bf 5.2 Scaling properties}
\bigskip
In statistical physics a dynamical system described by an order
parameter $\psi(\eta, x^i)$ is said to follow a scaling law if the
time evolution of its statistical properties depends only on a
time-dependent length scale $L(\eta)$, so that its equal time
auto-correlation function can be written in the form $$\langle
\psi(\eta,x^i)\psi(\eta,x'^i)\rangle =F(r/L(\eta)).\eqno(52)$$ (The
order parameter is supposed to be an homogeneous and isotropic random
field.)  In Fourier space the scaling law (52) translates as $$\langle
\hat\psi^*(\eta,k^i)\hat\psi(\eta,k'^i)\rangle
=\delta(k^i-k'^i)(2\pi)^{3/2}L^3(\eta)\hat F(kL(\eta)).\eqno(53)$$

In a cosmological context a natural length scale is the Hubble
radius~: $L(\eta)={\cal H}^{-1}=\eta$. Dimensionless ``order"
parameters $\psi(\eta, x^i)$ describing the network of topological
defects can be constructed using background quantities, such as
$\Theta^\mu_\nu(\eta,x^i)/\rho,$ where $\rho\propto1/\eta^4$ is the
background energy density. Numerical simulations of texture (see
e.g. [9] [10]) and cosmic string network (see e.g. [11]) evolution (as
well as qualitative arguments (see e.g. [7])) have
indicated that indeed these dimensionless random fields (but not the
random fields themselves) do obey, soon after the transition, a
scaling law (in fact, if they did not they would be either irrelevant
or catastrophic) with the Hubble radius as length scale.

We therefore have $$\langle
\Theta_{\mu\nu}(\eta,x^i)\Theta_{\mu\nu}(\eta,x'^i)\rangle =
{1\over\eta^4}F_{\mu\nu}(r/\eta)\eqno(54)$$ so that the power spectra
are of the form $$\langle
\hat\Theta^*_{\mu\nu}(\eta,k^i)\hat\Theta_{\mu\nu}(\eta,k'^i)\rangle
=\delta(k^i-k'^i)(2\pi)^{3/2}{1\over\eta}\hat
F_{\mu\nu}(k\eta).\eqno(55)$$ 

For statistically coherent sources described by (49) the
scaling property (54-55)  translates as
$$\hat S_a(\eta,k^i)={1\over\sqrt\eta}f_a(k\eta)\,e_A(k^i).\eqno(56)$$
What remains to be determined then is the behaviour of the ten functions
$f_a (k\eta)$, as well as the subsets $A$.

\bigskip
{\bf 5.3 Causality constraints}
\bigskip

The detailed structure of each realisation of
$\Theta_{\mu\nu}(\eta,x^i)$ is of course very complex. On scales less
than the correlation length (of the order of the Hubble radius at the
epoch of transition), $\Theta_{\mu\nu}(\eta,x^i)$ is almost zero since
the scalar field is there in a true vacuum state. In fact, it is non
zero only around the location of the topological defects, decaying
more or less quickly to zero away from them, depending on whether the
defects are local or global [2].  Numerical simulations have shown that,
as time passes, local cosmic strings interconnect and produce loops
which decay by emission of gravitational radiation, in such a way that
the energy distribution scales like the background density (see
previous \&). In the case of textures, previously uncorrelated regions
become causally connected and the scalar field tends to take the same
true vacuum value in each region, so that the number density of
textures decreases inversely to the horizon volume~; their energy is
also redshifted away so that, again, the energy density distribution
scales like the background density.

Now, even if each realisation of the phase transition produces a
network of defects whose typical length scale is, at all times,
greater than the horizon, their ensemble is, because the network
appeared at a definite time, that is for causality reasons, completely
uncorrelated on scales larger than the horizon. Therefore, as stressed
e.g. by Turok [17], the unequal time correlators in two points $P(\eta,x^i)$
and $P'(\eta',x^{\prime i})$ are strictly zero
if the past light-cones of $P$ and $P'$ do not intersect on the surface of
the phase transition~: 
$$\langle\Theta_{\mu\nu}(\eta,x^i)\Theta_{\lambda\rho}(\eta',x'^i)\rangle =0
\qquad\hbox{if}\qquad |\vec x-\vec x'|>\eta+\eta'.\eqno(57)$$

Property (57) translates in Fourier space into the fact that the power
spectra are white noise on super horizon scales (that is for
$k\eta\ll1$). Indeed, because the correlators (57) have compact
supports their Fourier transforms are analytic in $k^i$.

Let us first consider the spatial components of the correlators. They can
be written, in the small
$k$ limit, as 
$$\langle
\hat\Theta^*_{ij}(\eta,k^i)\hat\Theta_{kl}(\eta',k'^i)\rangle
=\delta(k^i-k'^i)\left[A\delta_{ij}{\delta}_{kl}+B\left({\delta}_{ik}
{\delta}_{jl}+{\delta}_{il}{\delta}_{jk}
\right)\right]+...,\eqno(58a)
$$
where the coefficients $A$ and $B$ are independent of $k^i$ (but may depend
on  $\eta$ and $\eta'$). The right hand side term was obtained by requiring 
a constant tensor (with respect to $k^i$) which respects the index symmetries
of the correlator (see Turok et al. [18]).
More generally, the correlators are of the form
$$\langle
\hat\Theta^*_{ij}(\eta,k^i)\hat\Theta_{kl}(\eta,k'^i)\rangle
=\delta(k^i-k'^i)\left(
\epsilon_t t_{ij}t_{kl}+2\epsilon_u u_{i(k}u_{l)j}\right),
\eqno(58b)$$
with 
$$
t_{ij}=t_0{\delta}_{ij}+t_1k_ik_j,\quad u_{ij}=u_0{\delta}_{ij}+u_1k_ik_j,
\eqno(58c)$$
where $t_0$, $t_1$, $u_0$ and $u_1$ are analytic functions of $k^2$
($\epsilon_t=\pm 1, \epsilon_u=\pm1$). This is the only decomposition
which is compatible with the homogeneity and isotropy of the
distribution.

In Fourier space,  any rank 2 symmetric tensor can be decomposed 
into scalar, vector and tensor parts, respectively~:
$$ 
\hat\Theta_{ij}=\hat\Theta^S_{ij}+\hat\Theta^V_{ij}+\hat\Theta^T_{ij}\eqno(59a)
$$
with
$$\hat\Theta^S_{ij}\equiv
\left({1\over 2}P_{ij}P^{kl}+L_i^kL_j^l\right)\hat\Theta_{kl}\eqno(59b)$$
$$\hat\Theta^V_{ij}\equiv
\left(P_i^kL_j^l+L_i^kP_j^l\right)\hat\Theta_{kl}\eqno(59c)$$
$$\hat\Theta^T_{ij}\equiv
\left(P_i^kP_j^l-{1\over 2}P_{ij}P^{kl}\right)\hat\Theta_{kl},\eqno(59d)
$$
where the complementary projection operators $P_{ij}$ and $L_{ij}$ 
are defined as
$$ P_{ij}={\delta}_{ij}-k_ik_j/k^2\quad,\quad L_{ij}=k_ik_j/k^2 .\eqno(60)$$
As a exercise, one can easily check that the pressure $P^s$ and anisotropic
stress $\Pi^s$ defined in (13) are simply given in Fourier space by 
the expressions
$$ \hat P^s={\delta}^{ij}\hat\Theta_{ij}/3, \quad 
  k^2\hat\Pi^s={1\over 2}\left(P^{ij}- 2L^{ij}\right)\hat\Theta_{ij}.$$

It can first be shown, using the decomposition (59) with the generic term 
(58b-c), that any correlator between scalar and vector, or vector and tensor, 
or tensor and scalar spatial quantities vanishes 
(this in fact is a consequence of the isotropy of the distribution). 

Let us then consider first 
the correlators between scalar spatial quantities. Using the above expressions,
one  finds easily 
$$\langle{\hat P}^{s*}\hat P^s\rangle=(A+2B/3), \quad 
\langle {\hat\Pi}^{s*} \hat\Pi^s\rangle=3Bk^{-4}, \quad
\langle {\hat P}^{s*}\hat\Pi^s\rangle={\cal O}(k^0),\eqno(61) 
$$ where $\langle{\hat P}^{s*}\hat P^s\rangle$ etc stand for
$\langle{\hat P}^{s*}(\eta,k^i)\hat P^s(\eta',k'^i)\rangle$ and where
it is understood that all the correlators are proportional to
$\delta(k^i-k'^i)$.

 The  correlators between the other scalar 
terms are obtained by introducing the correlators mixing time and spatial 
indices, namely
$$
\langle \hat\Theta^*_{00}\hat\Theta_{00}\rangle=
C, \quad 
\langle\hat\Theta^*_{00}\hat\Theta_{ij}\rangle=
D_0{\delta}_{ij}+D_1k_ik_j, \quad 
\langle\hat\Theta^*_{0i}\hat\Theta_{0j}\rangle=
E_0{\delta}_{ij}+E_1k_ik_j, \eqno(62)
$$
$$ 
\langle \hat\Theta^*_{00}\hat\Theta_{0i}\rangle=
-iFk_i, \quad \langle \hat\Theta^*_{0k}\hat\Theta_{ij}\rangle=
ik_k\left(G_0{\delta}_{ij}+G_1k_ik_j\right)+2iG_2k_{(i}\delta_{j)k},\eqno(63) 
$$ 
  the nine functions
($C,D_0,D_1,E_0,E_1,F,G_0,G_1 G_2$) being analytic in $k^2$. Performing the
scalar, vector tensor decomposition we obtain 
$$
\langle{\hat \rho}^{s*}\hat \rho^s\rangle=C,
\quad \langle{\hat \rho}^{s*}\hat P^s\rangle=D_0+D_1 k^2/3,
\quad \langle{\hat \rho}^{s*}\hat \Pi^s\rangle=-D_1,
\quad \langle{\hat v}^{s*}\hat v^s\rangle=E_0k^{-2}+E_1,\eqno(64)
$$
and
$$
\langle{\hat \rho}^{s*}\hat v^s\rangle=F,
 \quad \langle{\hat v}^{s*}\hat P^s\rangle=\left(G_0+G_2/3\right)+G_1
k^2/3,
\quad \langle{\hat v}^{s*}\hat \Pi^s\rangle=-G_1-2G_2/k^2.\eqno(65)
$$

Similarly, one finds for the nonvanishing vector correlators 
$$
\langle{\hat {\bar\Pi}}_i^{s*}{\hat {\bar\Pi}}_j^s\rangle=Bk^{-2}P_{ij},
\quad \langle{\hat {\bar v}}^{s*}_i{\hat {\bar v}}^s_j\rangle=E_0 P_{ij},
\quad\langle{\hat {\bar v}}^{s*}_i{\hat {\bar\Pi}}_j^s\rangle=G_2P_{ij}
\eqno(66)
$$
and for the tensor correlators
$$
\langle{\hat {\bar\Pi}}^{s*}_{ij}{\hat {\bar\Pi}}^s_{kl}\rangle=
B\left(P_{ik}P_{jl}+P_{il}P_{jk}-P_{ij}P_{kl}\right).\eqno(67)
$$ 

These results imply that, as a consequence of causality requirements, 
the quantities (56) describing the sources can be divided into three 
statistically independent subsets as defined in subsection 5.1 : 
a scalar subset including $\rho^s$, $P^s$, $\Pi^s$ 
and  $v^{s}$, characterised by a random variable $e_S(k^i)$; a disjoint, 
independent vector subset including 
 ${\bar v}^s_i$ and  
${\bar\Pi}_i^s$, characterised by $e_V(k^i)$ ; finally a tensor subset 
containing  ${\bar\Pi}^s_{ij}$ and characterised by $e_T(k^i)$.

\bigskip
{\bf 5.4 Conservation equations}
\bigskip
The last general property satisfied by  the sources is the conservation
equations. These equations can constrain the correlators obtained 
in the previous subsection. Indeed , if one multiplies eq (15d) by 
$\hat v^{s*}$ and then takes the correlator, one finds that the leading 
term of $E_0$ must decay as $\eta^{-4}$. In the following, we shall neglect 
this decaying term which turns out to be negligible with respect to the scaling
solution (see below) and thus take
$E_0\propto k^2$. As a consequence,
$$
\langle{\hat v}^{s*}\hat v^s\rangle={\cal O}(k^0),
\quad \langle{\hat {\bar v}}^{s*}_i{\hat {\bar v}}^s_j\rangle
={\cal O}(k^2)P_{ij}.\eqno(68)
$$

In order to maintain statistical coherence among the scalar subset of
variables, we shall now restrict our study to the case where the
dominant $k^{-4}$ dependence of the autocorrelator of $\hat\Pi^s$ (see eq. (61))
vanishes (otherwise, $\hat\Pi^s$ would go like $k^{-2}$ while
$\hat\rho^s$ and $\hat P^s$ go like $k^0$, which would imply, assuming
coherence, that the cross-correlators of $\hat\Pi^s$ with $\hat\rho^s$
or $\hat P^s$ would go like $k^{-2}$ in contradiction with (61) and
(64). Thus, taking into account the next order terms in the analytic
expansion, we shall assume

$$
\langle {\hat\Pi}^{s*} \hat\Pi^s\rangle={\cal O}(k^0), \quad
\langle{\hat {\bar\Pi}_i}^{s*}\hat {\bar\Pi}_j^s\rangle={\cal O}(k^2) P_{ij},
\quad
\langle{\hat {\bar\Pi}_{ij}^{s*}\hat {\bar\Pi}_{kl}^s}\rangle={\cal O}(k^4)
\left(P_{ik}P_{jl}+P_{il}P_{jk}-P_{ij}P_{kl}\right). \eqno(69)
$$

Combining our results with the scaling property (55-56) completely
determines on super-horizon scales the auto-correlators (48) or,
equivalently for our purposes, the random fields (49). Introducing the
scalar, vector and tensor decomposition performed in section 2, the
random fields describing statistically coherent sources that scale,
satisfy the causality and conservation constraints finally are~:
$$\eqalign{
\hat\rho^s&={1\over\sqrt\eta}f_1({k\eta}) e_S(k^i),\quad \hat
P^s={1\over\sqrt\eta}f_2({k\eta})e_S(k^i),\cr
\hat v^s&=-\sqrt\eta f_3(k\eta)e_S(k^i),\quad\hat
\Pi^s=\eta^{3/2}f_4(k\eta)e_S(k^i),\cr}\eqno(70)$$
$$\hat {\bar v}_i^s= k\sqrt{\eta}\bar g_{1i}({k\eta})e_V(k^i), \quad\hat {
\bar\Pi}_i^s=k\eta^{3/2}\bar g_{2i}(k\eta)e_V(k^i),
\quad \hat {\bar\Pi}^s_{ij}=k^2\eta^{3/2}\bar
 h_{ij}({k\eta})e_T(k^i)\eqno(71)$$ 
where the ten\footnote{$^1$}{In fact there
are six such functions only~: see ``Cosmic microwave background

anisotropies seeded by coherent topological defects~: a
semi-analytical approach''

 by J.P. Uzan, N. Deruelle and A.Riazuelo}
functions $f_a(k\eta), \bar g_{ai}(k\eta),\bar h_{ij}(k\eta)$ tend to
constants on super horizon scales. On smaller scales ($k\eta\gg1$)
they tend to zero (in an oscillatory fashion) at a rate which depends
on whether the defects are local or global and which can be
determined precisely solely by means of numerical
simulations. However, since our purpose is to set initial conditions
at a time when all relevant scales are outside the horizon, we shall
not need to know these functions for $k\eta>1$.

\bigskip
\bigskip
\noindent
{\cs 6. The long wavelength solution after the transition}
\bigskip
We shall now construct the long-wavelength solutions, that is valid as
long as $k\eta\ll1$, of the equations of motion written in Fourier
space for the scalar (eq (15)), the vector (eq (23)) and the tensor (eq
(26)) perturbations, subject to the matching conditions (41) at
$\eta=\eta_{PT}$.  The source terms will be supposed to be given by Eq
(70-71) where the functions $ f_a(k\eta),
\bar g_{ai}(k\eta), \bar h_{ij}(k\eta)$ can be expanded 
in power series of $k\eta$ and tend to
constants when $k\eta\to0$~: $$ f_a(k\eta)\to A_a\quad a={1,2,...4},
\quad \bar g_{ai}\to \bar B_{ai}\quad a={1,2}, \quad \bar
h_{ij}\to\bar C_{ij}\quad\hbox{when}\quad k\eta\to0.\eqno(72)$$ The
long wavelength
description of the topological defects is thus reduced to the giving
of ten constants.

\bigskip
{\bf 6.1 Scalar modes}
\bigskip
The four scalar source terms $(\hat\rho^s,\hat P^s,\hat v^s,\hat
\Pi^s)$ are subject to the two constraints (15c-d). In the
long-wavelength limit (72), these constraints read~: $$ A_1=-6
A_2\qquad,\qquad A_3={2\over5}A_2.\eqno(73)$$ The general long
wavelength solution of Eq (15h) for $\Psi$ is (taking into account
(73)) $$\hat
\Psi=\Psi_0+{\Psi_1\over\eta^3}+{2\over9}\kappa\eta^{3/2}(A_4+A_2)\eqno(74)$$
where $\Psi_0$ and $\Psi_1$ are two constants of integration and
where, from now on, we omit writing the normalised random field
$e_S(k^i)$ to which each scalar perturbation is proportional to.  Eq (15e-f-g)
then give the other perturbations $$\eqalign{\hat
\Phi=\Psi_0+{\Psi_1\over\eta^3}+{1\over9}\kappa\eta^{3/2}(2A_2-7A_4)
\quad,
\quad\hat \delta^\natural={8\over5}\kappa\eta^{3/2}A_2\cr
\hat
v^\natural=-{\eta\over2}\Psi_0+{\Psi_1\over\eta^2}
-{2\kappa\eta^{5/2}\over45}(4A_2-5A_4).
\cr}\eqno(75)$$
Equation (12) then gives~: $$\hat
\delta^\flat=-6\Psi_0\qquad,\qquad\hat
\delta^{\sharp}=-2\Psi_0+{4\Psi_1\over\eta^3}+ {8\over9}\kappa
\eta^{3/2}(A_2+A_4).\eqno(76)$$ (One can check that the Bianchi
identities (15a-b) are identically satisfied by the solution (74-75).)
Finally the constants $\Psi_0$ and $\Psi_1$ are determined by the
matching conditions (41)~: $$\Psi_0=-{1\over3}\kappa\eta^{3/2}_{PT}A_2
\quad,\quad{\Psi_1\over\eta^3_{PT}}={1\over9}\kappa\eta^{3/2}_{PT}(-2A_4+A_2).
\eqno(77)$$
One therefore sees that deep in the radiation era but long after the
transition, when $\eta\gg\eta_{PT}$, all terms proportional to $\Psi_0$
and $\Psi_1$ in the expressions for $\Psi,\Phi,\delta^\natural,
v^\natural$ and $\delta^\sharp$ can be neglected. We shall comment on
$\delta^\flat$ in the paragraph on compensation.

\bigskip
{\bf 6.2 Vector and tensor modes}
\bigskip

The two vector source terms ($\hat {\bar v}^s_i,\hat {\bar\Pi}^s_i$),
given by (71-72) are constrained by the conservation equation
(23b). In the long wavelength limit this leads to ${\bar
B}_{1i}=0$. At next order one gets 
$$\bar v^s_i={2\over9}\eta^{5/2}k^3\bar B_{2i},\eqno(78)$$ 
where, again, the normalised random field $e_V(k^i)$ to which each vector
perturbation is proportional has been omitted. The general long
wavelength solution of equation (23d) is
$$\hat{\bar\Phi}_i={\bar{\Phi}_{0i} \over\eta^2}+{4\over 9}\kappa
k\bar{B}_{2i}\, \eta^{5/2},\eqno(79)$$ where $\Phi_{0i}$ is a constant
of integration which is determined by the matching condition (41)~: $$
{\bar{\Phi}_{0i}\over\eta_{PT}^2}=-{4\over9}\kappa
k\bar{B}_{2i}\,\eta_{PT}^{5/2}.  \eqno(80)$$ Finally (23c) yields $$\hat
{\bar v}_i^{\sharp}={k^2\over8}\bar\Phi_{0i}\eqno(81)$$ (so that the
Bianchi identity (23a) is satisfied).  \vskip 0.5cm The general
solution of equation (26) for the tensorial modes, for a source given
by (71)(72), is$$ \hat{\bar E}_{kl}=\left[\bar A_{kl}+{\bar
B_{kl}\over\eta}+{8\over63}\kappa k^2 \bar
C_{kl}\,\eta^{7/2}\right] e_T(k^i),\eqno(82)$$ where $\bar A_{kl}$ 
and $\bar B_{kl}$
are two integration constants determined by the matching conditions
(41)~: $$ \bar A_{kl}=-{4\over7}\kappa k^2\bar
C_{kl}\,\eta_{PT}^{7/2}\quad,\quad {\bar
B_{kl}\over\eta_{PT}}={4\over9}\kappa k^2\bar
C_{kl}\,\eta_{PT}^{7/2}.\eqno(83)$$ One sees again that deep in the
radiation era but long after the transition, all terms proportional to
the integration constants
 can be neglected. 

\bigskip
{\bf 6.3 Initial conditions for Cold Dark Matter}
\bigskip

We assumed that at the epoch of the transition the matter content of
the universe was a radiation fluid consisting of photons, neutrinos
and highly relativistic particles, some coupled to the photons, some
not (being ``WIMPS", see e.g. [19]).  Thus the density contrasts that we have
introduced ($\delta^\natural,\delta^{\sharp},\delta^\flat$) are density
contrasts of the radiation fluid.  Now as the universe expands and its
temperature drops, the WIMPS become non relativistic and turn into 
cold dark matter (CDM).
 In this paragraph we discuss the evolution of the perturbations
of this pressureless component of the cosmic fluid, when the universe
is still radiation dominated.

The background energy momentum tensor of CDM simply is~: ${\cal
T}_{\mu\nu}=\rho_cu_\mu u_\nu$ with $u_0=-a,u_i=0$ ($a$ being the
scale factor) and where $\rho_c$ is the energy density. Since CDM, by
definition, interacts only gravitationaly with the radiation fluid,
${\cal T}_{\mu\nu}$ is conserved so that~$ \rho'_c/\rho_c=-3/\eta$.

The perturbations of ${\cal T}_{\mu\nu}$ can be decomposed 
into scalar vector
and tensor components which can be written as
$$\delta {\cal T}^S_{00}=a^2\rho_c(\delta_c+2A)\quad,\quad\delta
{\cal T}^S_{0i}=-a^2\rho_c\partial_i(B+v_c)\quad,\quad
\delta
{\cal T}^V_{0i}=-a^2\rho_c(\bar B_i+\bar v_c^i)\eqno(84)$$ (the other
components being zero) where $\delta_c,v_c$ and $\bar v_c^i$ are the
density contrast and the velocity components of the perturbations of
the CDM fluid. In the coordinate transformation (5)
$\delta_c\to\delta_c -3T/\eta$. As for $v_c$ and $\bar v_c^i$ they
transform like the radiation velocities. Various gauge invariant
perturbations can therefore be built, such as~:
$$\delta_c^\flat=\delta_c+3C\quad,\quad
v_c^\natural=v_c+E'\quad,\quad \bar v_c^{{\sharp}\,i}=\bar v_c^i+\bar
B^i.\eqno(85)$$ Other gauge invariant density contrasts can be
introduced, e.g.  $$\delta^\natural_c=\delta_c-{3\over\eta}(v+B)
\qquad\hbox{or}\qquad\delta_c^{\sharp}=\delta_c-{3\over\eta}(B-E').\eqno(86)$$
Note that $\delta_c^\natural$ is the density contrast in the gauge where
the radiation fluid is at rest.

Since radiation still dominates, the density contrast and velocity
perturbations of the total fluid are almost equal to the radiation
density constrast $\delta$ and the radiation velocities $v$ and $\bar
v^i$, so that the Einstein equations (15e-h), (23c-d) and (26) remain
unchanged. Therefore the solution given in the previous paragraphs for
the metric perturbations as well as for the (dominant) radiation
component still holds even after the WIMPS have become non
relativistic.

As for the evolution of the  perturbations of the CDM it is governed
 by the conservation
equations of its energy-momentum tensor, that is (in Fourier space)~:
$$\hat\delta_c^{\flat'}=k^2\hat v_c^\natural\qquad,\qquad\hat
v_c^{\natural'}+{1\over\eta}\hat v_c^\natural=-\hat\Phi\eqno(87)$$
$$\hat v_c^{{\sharp}i'}+{1\over\eta}\hat v_c^{{\sharp}i'}=0,\eqno(88)$$
where $\hat\Phi$ is given by (75). The general solution of this system is, 
on super-horizon
scales
$$\hat\delta_c^\flat=\delta_0\quad,\quad
\hat v_c^\natural={v_0\over\eta}-{\Psi_0\over2}\eta+{\Psi_1\over\eta^2}-
{2\over63}(2A_2-7A_4)\kappa\eta^{5/2}\eqno(89)$$ $$\hat {\bar
v}_c^{{\sharp}i}={\bar v_0^i\over\eta},\eqno(90)$$ where $\delta_0$, $v_0$
and $\bar v_0^i$ are integration constants, which contrarily to
$\Psi_0$ and $\Psi_1$ are not given by the matching conditions at the
time of the transition.  Rather they should be deduced from a detailed
analysis of the formation of the WIMPS. As for the other density
contrasts they are given by 
$$\hat\delta_c^\natural=\delta_0+{9\over2}\Psi_0+
{6\over5}A_2\kappa\eta^{3/2}\quad,\quad
\hat\delta_c^{{\sharp}}=\delta_0+3\Psi_0+{3\Psi_1\over\eta^3}+{2\over3}
(A_2+A_4)\kappa\eta^{3/2}.\eqno(91)$$

Once more one sees that deep in the radiation era but long after the
transition, all terms proportional to $\Psi_0$ and $\Psi_1$ in the
expressions for $\hat \delta_c^\natural,\hat v_c^\natural$ and $\hat
\delta_c^{\sharp}$ can be neglected. As for the terms proportional to the
unknown constants $v_0$ and $\bar v_0^i$ they are decaying modes.  The
only new constant of potential dynamical relevance is therefore
$\delta_0$.

\vskip0.5cm
Let us now compare the density contrasts of CDM and radiation. The
isocurvature perturbation is the gauge invariant quantity $\hat S\equiv\hat
\delta_c-{3\over4}\hat \delta$. From (75) and (91) 
we have that it is given, at lowest order in $k$, by
$$
\hat S=\delta_0+{9\over2}\Psi_0\eqno{(92)}
$$
We note that in $\hat S$, just like in the density contrasts $\hat\delta^\flat$
and $\hat\delta^\flat_c$, the terms of the type $\kappa A_i\eta^{3/2}$
are absent. We must therefore compute them at next order in
$k$. Injecting the asymptotic behaviours (75) and (89)
for the velocity perturbations into (15a) and (87),
we obtain~:
$$
\hat S=\delta_0+{9\over2}\Psi_0+{8\over245}\kappa A_2 (k\eta)^2
\eta^{3/2}.\eqno{(93)}
$$
In the comoving gauge (in which the radiation fluid is at rest), the ratio
of isocurvature to total perturbations is~:
$$
{\hat S^\natural\over\hat
\delta^\natural}\to {1\over49}(k\eta)^2,\eqno{(94)}
$$
and does not depend on the detailed structure
of the defects [as long as, of course, they can be represented by the
random fields (70-71)].

\bigskip
\noindent
{\cs 7. Comments and dicussion}
\bigskip

In this Section we briefly compare our results to the initial
conditions currently used in the literature.
\bigskip
{\bf 7.1 ``Natural" initial conditions}
\bigskip
In [22-23], Durrer, Sakellariadou and Gangui studied the
perturbations of the radiation and CDM components of the cosmic fluid
seeded by the sources (70). They chose ``natural" initial
conditions, which amounts to setting $\Psi_0=\Psi_1=0$ in (74-76) (they
considered the scalar perturbations only). Our work justifies their
choice since, as we saw, the homogeneous modes (in the sense of
differential equations) of the evolution equations become negligible
fairly soon after the phase transition. Their solution however differs
from ours~: indeed they only solve the conservation equations for the
radiation and CDM fluids (that is Eq. (15 a-b) and (87)) together with
the Einstein  equations (15 e-f). They therefore ignore the
conservation equations for the sources, that is Eq. (15 a-b), arguing
that they do not hold for ``incoherent" sources.  Their
solution hence depends on three constants, $A_1, A_3$ and $A_4$,
instead of two. However it is clear that the conservation equations
for the sources (or, equivalently, the Einstein equations (15
g-h)) {\it must} be satisfied, whether the sources are described by
ordinary functions or random fields, and that $A_3$ in [22-23] should
be set equal to $-{1\over15}A_1$.

\bigskip
{\bf 7.2 Fixing the initial conditions by means of a ``pseudo-tensor"}
\bigskip

The evolution of the perturbations in synchronous gauges where
$A=B=\bar B_i=0$, which are used by a number of authors, in particular
Pen, Spergel and Turok [9], can be straightforwardly obtained from the 
definition
(10-11) (85-86) of the gauge invariant quantities we introduced and
their explicit expressions (74-77) (89-91). For example~: $$\hat
\delta_c^{syn}=\delta_0+{9\over2}\Psi_0-
{3e_0\over\eta^2}+{6\over7}\kappa\eta^{3/2}\,A_2\quad\hbox{and}\quad
\hat \delta^{syn}=-{4e_0\over\eta^2}+{8\over7}\kappa\eta^{3/2}\,A_2\eqno(95)$$
where $e_0$ is an extra constant of integration linked to the fact
that synchronous gauges are not completely fixed.

Now the solution (95) can of course also be obtained by a direct
integration of the perturbation equations (15) written in the
synchronous gauge. It is an easy exercice to show from Eq (15) and
(87), and (70) (72), that the synchronous gauge density constrasts for
the CDM and the radiation fluid satisfy, on super-horizon scales, the
following equations $$\hat \delta^{syn\prime}_c={3\over4}\hat
\delta^{syn\prime}\quad,\quad\hat
\delta^{syn\prime\prime}+{\hat \delta^{syn\prime}\over\eta}-{4\hat
\delta^{syn}\over\eta^2}=-2{A_2\over\sqrt\eta}
\eqno(96)$$
the general solutions of which are $$\hat \delta_c^{syn}={3\over4}\hat
\delta^{syn}+D_0\quad\hbox{and}\quad\hat \delta^{syn}=
D_1\eta^2+{D_2\over\eta^2}+{8\over7}A_2\kappa\eta^{3/2}
\eqno(97)$$
where $D_0,D_1,D_2$ are three integration constants. After proper
identification of the integration constants one sees that (95) and
(97) are indeed equivalent, up however to the growing mode proportional to
$D_1\eta^2$. In order to eliminate this growing mode, a conserved
``pseudo-tensor" $\tau_{\mu\nu}$, i.e. such that
$\partial_\mu\tau^\mu_\nu=0$, was built by Turok et al. [9] [16].  On
superhorizon scales the $00$ component of this pseudo-tensor is
$$\kappa \tau_{00}={3\hat\delta^{syn}\over\eta^2}-{6\hat
C'\over\eta}+{\kappa A_1\over\sqrt\eta}\eqno(98)$$ and is, because
$\tau_{\mu\nu}$ is conserved, constant. Since before the phase
tansition it was zero, it must be zero at all times~:
$$\tau_{00}=0.\eqno(99)$$ Now, from (15a) and (10), $\hat C'=-\hat
\delta'/4$. Using (97) one then indeed sees that the condition
$\tau_{00}=0$ is equivalent to imposing $D_1=0$.

Introducing a pseudo-tensor is therefore a way to insure that the
solution (97) is a solution of Einstein's equations. In fact condition
(99) is nothing but the relativistic Poisson equation written in
synchronous gauge.  It is {\it not} a way to choose the four true
integration constants of the general solution of Einstein's equations
(15) and (87), that is $\Psi_0$ and $\Psi_1$, $\delta_0$ and
$v_0$. These constants are fixed by Turok et al. [9] [17] by choosing the
solution of (87) ($\hat\delta_c^{syn\prime}+3\hat C'=0$) to be
$\hat\delta_c^{syn}=-3\hat C$ on the grounds that before the
transition all perturbations were zero, and by choosing ``adiabatic"
perturbations, that is imposing $\hat S^{syn}=0$. This amounts to
choosing $\Psi_0=\delta_0=0$, which contradicts (77) but, as we
saw, this does not really matter since the solution at late time is not
sensitive to the matching conditions.

\bigskip
{\bf 7.3  ``Integral constraints" and compensation}
\bigskip

Trashen, in [24], introduced a new way to extract from Einstein's
equations integral equalities which relate the total matter
perturbations on a Roberston-Walker background. They are defined in
synchronous gauge and read $$\eqalign{\int_\Sigma(\delta T^0_0-{\cal
H}\delta T^0_kx^k)d^3x&=
\int_{\partial\Sigma}B^ldS_l\cr
\int_\Sigma\left(\delta T^i_0+{\cal
H}\delta T^0_l\left[{1\over2}\delta^{il}x^px_p-x^lx^i\right]\right)d^3x&=
\int_{\partial\Sigma}B^{li}dS_l\cr}
\eqno(100)$$
where $\delta T^\mu_\nu$ is the perturbation of the total
energy-momentum tensor of the defects and the cosmic fluid, where the
quantities $B^l$ and $B^{li}$ are linear in the metric perturbations,
and where $\Sigma$ is a comoving three-volume, $\partial\Sigma$ being
its boundary.

These integral equalities can be interpreted as defining the energy of
the perturbations (see [25]) and have been much invoked in the
literature on cosmological perturbations seeded by topological defects
(see e.g.  [8-9] [26-27]). To understand better their role we shall restrict
our attention to scalar perturbations and shall rewrite them, in the
context here considered of a radiation dominated universe, as (see (4)
(9) (10-11) (13))

$$\int_\Sigma{\cal U}d^3x=\int_{\partial\Sigma}C^ldS_l\quad,\quad
\int_\Sigma{\cal U}x^i
d^3x=\int_{\partial\Sigma}C^{li}dS_l\quad\hbox{with}\quad{\cal U}\equiv
{3\delta^\natural\over\eta^2}
+3\kappa\left[\rho^s-{3\over\eta}v^s\right]
\eqno(101)$$
where $C^l$ and $C^{li}$ are different from $B^l$ and $B^{li}$ because
integration by parts were performed. Written under the form (101)
Trashen's equalities relate gauge invariant quantities.

As always with conservation laws in general relativity, eq (101) give
some information about the solution of Einstein's equations without
having to actually solve them. In the case at hand they tell the
following~: if the sources of the perturbations are localised within
one horizon volume or if, as is the case with topological defects,
they are randomly distributed on scales larger than the horizon, then
the surface integrals on the right-hand side of (101) vanish for
$\Sigma$ larger than a horizon volume. The left-hand sides are hence
also zero and, for scales larger than the horizon, the integrands
being almost constant must also vanish. Now the equation ${\cal U}=0$
is just Eq (15f) on super-horizon scales, that is nothing but another
rewriting of the relativistic Poisson equation.

Trashen's equalities were also interpreted in terms of ``compensation"
[8-9] [21] [27-28].  The idea here is that if the surface integrals in (101)
vanish, then, as shown by Abbot and Trashen [29], the Fourier
transform of ${\cal U}$ must be of order $k^2$. Now, as we have seen,
the Fourier transforms of the source functions, e.g. $\rho^s$ and
$v^s$, are white noise i.e. of order $k^0$.  Consequently,
$\delta^\natural$ must also be white noise, perfectly correlated to
the sources so as to ``compensate" them.  Again this is a rephrasing
of the Poisson equation.

Finally, the fact that $\delta^\flat$ is much smaller than $\Psi$
deep in the radiation era (see eq (74-76)) has also been interpreted by
Durrer and Sakellariadou [23] in terms of ``compensation'' but it is
perhaps simpler to say that this property follows from the continuity
equation (15a).

In conclusion, we believe that the somewhat heavy terminology used in
the literature of ``pseudo-tensor'', ``integral constraints'' or
``compensation'' is not really required if one straightforwardly
solves Einstein's equations.

\bigskip
\noindent
{\cs Acknowledgments}
\bigskip
We are very grateful to Ruth Durrer and Mairi Sakellariadou for
numerous and enjoyable discussions. We warmly thank Neil Turok for
enlightening explanations and for pointing out a mistake in eq. (65).
We thank R. Brandenberger for interesting remarks as well as Robert
Caldwell and Joao Magueijo for commenting their work.
\bigskip
\noindent
{\cs References}
\bigskip

[1] P.J. Steinhardt, ``Cosmology at the Crossroads'', to appear in the
{\it Proceedings of the Snowmass Workshop on Particle Astrophysics and
Cosmology}, E. Kolb and R. Peccei eds., 1995; astro-ph/9502024.

[2] T.W.B. Kibble, Phys. Rep. {\bf 67}, 183, 1980; A. Vilenkin and
E.P.S. Shellard, ``{\it Cosmic strings and other Topological
Defects}'', Cambridge University Press, 1994.

[3] T.W.B. Kibble, J. Phys {\bf A9},1387, 1976.

[4] J.M. Bardeen, Phys. Rev. {\bf D22}, 1882, 1981.

[5] H. Kodama and M. Sasaki, Prog. Theor. Phys. Suppl., 78, 1984.

[6] V.F. Mukhanov, H.A. Feldman and R.H. Brandenberger,
Phys. Rep. {\bf 215}, 203, 1992.

[7] R. Durrer, Fund. of Cosmic Physics {\bf 15}, 209, 1994.

[8] S. Veeraraghavan and A. Stebbins, Astrophys. J. {\bf 365}, 367,
1990.

[9] U-L. Pen, D.N. Spergel and N. Turok, Phys. Rev. {\bf D49}, 692,
1994.

[10] R. Durrer and Z.H. Zhou, Phys. Rev. {\bf D53}, 5384, 1996.

[11] B. Allen, R. R. Caldwell, S. Dodelson, L. Knox, E. P. S. Shellard
and A. Stebbins, astro-ph/9704160; B. Allen, R. R. Caldwell,
E. P. S. Shellard, A. Stebbins and S. Veeraraghavan, astro-ph/9609038.

[12] A. Guth, Phys. Rev. {\bf 23}, 347, 1981.

[13] A. Lichn\'erowicz, {\it Th\'eories relativistes de la gravitation
et de l'\'electromagn\'etisme}, ed. Masson, Paris 1955.

[14] N. Deruelle and V.F. Mukhanov, Phys. Rev. {\bf D52}, 5549, 1995.

[15] N. Turok, astro-ph/9704165.

[16] J. Magueijo, A. Albrecht, P. Ferreira and D. Coulson,
astro-ph/9605047.

[17] N. Turok, Phys. Rev. Lett {\bf 77}, 4138, 1996; N. Turok,
Phys. Rev. {\bf D54}, 3686, 1996.

[18] N. Turok, U-L. Pen and U. Seljak, astro-ph/9706250

[19] P.J.E. Peebles, ``{\it Principles of Physical Cosmology}'',
Princeton University Press, 1993.
E.W. Kolb and M.S. Turner, ``{\it The Early Universe}'',
Addison-Wesley, 1990 (chap. 6,16-18).

[20] W. Hu, D.N. Spergel and M. White, astro-ph/9605193.

[21] C. Cheung and J. Magueijo, astro-ph/9702041.

[22] R. Durrer, A. Gangui and M. Sakellariadou, Phys. Rev. Lett. {\bf
76}, 579, 1996.

[23] R. Durrer and M. Sakellariadou, astro-ph/9702028; to appear in
Phys. Rev. D

[24] J. Traschen, Phys. Rev {\bf D31}, 283, 1985.

[25] N. Deruelle, J. Katz and J-P. Uzan, Class. Quant. Grav. {\bf 14},
421, 1997.

[26] J. Traschen, Phys. Rev. {\bf D29}, 1563, 1984; J. Traschen and
D. Eardley, Phys. Rev. {\bf D34}, 1665, 1986.

[27] J. Magueijo, Phys. Rev. {\bf D46}, 1368 and 3360, 1992.

[28] A. Albrecht and A. Stebbins, Phys. Rev. Lett. {\bf 68}, 2121,
1992; A. Albrecht and A. Stebbins, Phys. Rev. Lett. {\bf 69}, 2615,
1992.

[29] L.F. Abott and J. Traschen, Astrophys. J. {\bf 302}, 39, 1986.

\end